\documentclass[twoside,amsmath,amssymb,nofootinbib,superscriptaddress,notitlepage]{revtex4-1}
\usepackage{graphicx,subfigure}
\usepackage{dcolumn,marvosym}
\usepackage{float,hyperref}
\usepackage{empheq,cleveref}
\usepackage{verbatim}
\usepackage{color}
\usepackage{enumitem}
\usepackage{amsmath}
\usepackage{amsfonts}
\usepackage{graphicx}
\usepackage{bookmark}
\begin{document}

\title{The melting of charmonium in a magnetic field from an effective AdS/QCD model}
\author{David Dudal}
\email{david.dudal@kuleuven-kulak.be}
\affiliation{KU Leuven Campus Kortrijk - KULAK, Department of Physics, Etienne Sabbelaan 53, 8500 Kortrijk, Belgium}
\affiliation{Ghent University, Department of Physics and Astronomy, Krijgslaan 281-S9, 9000 Gent, Belgium}
\author{Thomas G.~Mertens}
\email{thomas.mertens@ugent.be}
\affiliation{Ghent University, Department of Physics and Astronomy, Krijgslaan 281-S9, 9000 Gent, Belgium}

\begin{abstract}
We study the influence of a background magnetic field on the melting of the $J/\psi$ vector meson by introducing a Born-Infeld modification of the soft-wall model. Out of the three polarizations of the massive vector meson, we find that the longitudinal one (parallel to the applied magnetic field) melts only at an even higher temperature than the deconfinement temperature, whereas the two transverse polarizations melt at a lower temperature than in the absence of a magnetic field. We also conduct a preliminary investigation of the effect of the magnetic field on the heavy quark diffusion coefficient, showing an increased diffusion constant for the longitudinal polarization with respect to the transverse polarizations.
\end{abstract}

\maketitle
\bookmarksetup{startatroot}

\section{Introduction}
The quark-gluon plasma is a short-lived exotic state of QCD matter, in earth circumstances only present in controlled laboratory environments, that is after a heavy ion collision (e.g.~using Pb- or Au-ions). The ruling experiments are RHIC and the LHC Alice facility. The state of deconfined quark-gluon matter is exotic, not only because it is hard to create but because its properties are not always as expected from a standard plasma \cite{Shuryak:2008eq}. Despite the ultrahigh temperature at which the plasma is existing ($\sim 10^{12}~K$), the plasma degrees of freedom are still strongly coupled, making a perturbative analysis of the  physics virtually impossible. This strong coupling nature of the plasma can be intuitively understood from the observation that the relevant temperature ---when reexpressed in more appropriate MeV units--- becomes of the order of the fundamental QCD scale $\Lambda_{QCD}$, at which the QCD-interaction becomes strong.

A potential extra ingredient to further complicate matters was identified a few years ago \cite{Kharzeev:2007jp,Skokov:2009qp,Bzdak:2011yy}. It was discussed how a noncentral heavy ion collision can produce very strong (short-lived) magnetic fields as well ($\sim 10^{16}~$Tesla). Such strong magnetic ($B$) fields can couple directly to the electrically charged quarks and, via the quarks, also indirectly to the gluons. As this magnetic field is a collective effect, we can treat it as a classical external field and due to its size, one might expect considerable effects onto the QCD dynamics. The recent review \cite{Kharzeev:2013jha} contains a pleiad of new strange QCD phenomena directly linked to the presence of a strong magnetic background. We point out that most studies restrict to a constant (in time and position) magnetic field of the form $\mathbf{B}=B\mathbf{e}_z$. There have been given (model dependent) estimates of the magnetic field showing that during the initial stages after the collision during which the plasma exists ($\sim 1-10$~fm), it is nearly constant \cite{Tuchin:2013apa}, hence this approximation is a decent one that allows explicit analyses, at least to some extent.

A particular riddle in magnetized QCD is the effect the magnetic field might have on the deconfinement transition. An approach based on the linear sigma model coupled to the Polyakov loop \cite{Mizher:2010zb} predicted an increasing $T_c$ in terms of $B$, a result supported by an independent study using the Polyakov-Nambu-Jona-Lasinio model \cite{Fukushima:2010fe,Gatto:2010pt,Gatto:2010qs}, while earlier a nonlinear sigma model analysis predicted a decrease \cite{Agasian:2008tb}. Quenched lattice QCD studies on one hand seemed to support a (slight) decrease in $T_c$ \cite{D'Elia:2010nq} at first, but the situation got more or less settled when unquenched results (full quark dynamics) became available \cite{Bali:2011qj}; a direct study of the Polyakov loop or (strange) quark susceptibility displayed a decreasing critical temperature for the confinement-deconfinement transition. Let us also refer to \cite{Ilgenfritz:2012fw,Ilgenfritz:2013ara} for two-color QCD studies.

By now, also some analytical approximations were worked out which were capable of explaining a decreasing $T_c$, see e.g.~\cite{Fraga:2012fs,Fraga:2012ev}. It is perhaps interesting to point out here that, frequently, the discussion about the deconfinement transition in a magnetic field is coupled to that of the chiral transition. Since quarks couple directly to the magnetic field, one could expect an even stronger response of the chiral transition to the $\mathbf{B}$-field. The predictions of the papers \cite{Mizher:2010zb,Gatto:2010pt,Gatto:2010qs,Fukushima:2010fe,Bali:2011qj,D'Elia:2010nq,Ilgenfritz:2012fw,Ilgenfritz:2013ara} ranged from a split or no split between the two transition temperatures, with the temperatures rising or going down with $B$. The already mentioned lattice QCD study \cite{Bali:2011qj} in fact was most focused on disfavoring this rising chiral transition temperature, rather showing the (unexpected) opposite behavior: the chiral transition temperature also goes down with $B$. This triggered a lot of attention, see the already cited papers and also \cite{Kharzeev:2013jha}.

In the current note, we will exclusively be interested in the deconfinement transition in the presence of a magnetic field. We will employ the contemporary gauge-gravity duality approach \cite{Maldacena:1997re} applied to QCD-like theories. Although the original AdS/CFT duality was not really formulated for a gauge theory like QCD, it is safe to say that the original AdS/CFT insights inspired quite some activity in gauge theories. Some of these can be called cousins of QCD (see e.g.~\cite{Sakai:2004cn,Sakai:2005yt,Kruczenski:2003be,Kruczenski:2003uq,Erdmenger:2007cm} for some benchmark papers) while others are effective dual descriptions onto which the AdS/CFT technology is unleashed, see e.g.~\cite{Erlich:2005qh,Karch:2006pv}. We will in particular take a look at a specific ``foreteller'' of the deconfinement transition in a specific AdS/QCD setting. We recall here that a famous smoking gun signal of the formation of the quark-gluon plasma is $J/\psi$-suppression, as proposed in \cite{Matsui:1986dk}, a charm-anticharm bound state. Since then, charmonium properties have received a widespread attention. In particular, let us refer to the lattice paper \cite{Asakawa:2003re} in which by means of the numerical study of lattice correlation functions, it was found that charmonia can survive the deconfinement transition at critical temperature $T_c$ and continue to exist up to $T\sim 1.6~T_c$. We recall here the critical temperature is usually defined and extracted (approximately) via the Polyakov loop \cite{Greensite:2003bk}, a quantity that has no direct connection to observable particle physics. At the level of observable features of the quark-gluon plasma, a study of charmonium melting (dissociation) at high temperatures might be a more practical way to look at the deconfinement transition.

We will thus be interested in the melting of charmonia in a holographic model that we believe to capture some essential features of magnetized QCD. We are not the first to investigate meson melting in AdS/QCD \cite{Peeters:2006iu,Fujita:2009wc,Fujita:2009ca,Ishii:2014paa,Ali-Akbari:2014gia,Ali-Akbari:2013txa,AliAkbari:2012vt}, but to our knowledge, determining how the magnetic field influences the melting temperature has never been considered. The hope is that this can learn us something about the deconfinement transition in a magnetic field. Usually, in the dual gravitational approach, the quarks are added in a probe brane approximation, following the seminal proposal \cite{Karch:2002sh}. As a result, the backreaction of the quarks onto the gluonic (confining or deconfining) dual geometric background is not taken into account. Since it is exactly this background that determines the confinement-deconfinement transition (via a Hawking-Page like transition \cite{Herzog:2006ra,Aharony:2006da}), there is no direct quark effect on the critical temperature $T_c$. A fortiori, this means that there is no magnetic field effect either on $T_c$, since $B$ can only couple to the uncharged glue sector via the charged quarks \cite{Johnson:2008vna,Callebaut:2013ria}. One possible way to make the transition nonetheless $B$-dependent was discussed in \cite{Ballon-Bayona:2013cta}, by taking into clever account the pressure of the quark probe branes into the derivation of the critical temperature, while motivating that a complete analysis of the backreaction would be of higher order in the number of flavors over the number of colors. The result was a decrease in the deconfinement temperature.  In the current work, we will investigate how the melting of charmonia is influenced by a magnetic field. The complication arising is the neutral nature of a $c\overline c$ bound state, so we will need to couple the magnetic field to the substructure of the bound state.

The paper is organized as follows. In Sections II and III, we introduce a soft wall model that allows to incorporate to some extent the coupling of the magnetic field to the charged quark constituents in the $J/\psi$ charmonium. The equations of motions are analyzed in Section IV, followed by a numerical evaluation of the spectral functions in Section V. Subsequently, we compare our results to lattice or other models' data. We also offer an exploratory study of the heavy quark diffusion constant in Section VII. Finally, conclusions and an outlook to future research are presented in Section VIII.

\section{The soft wall model and its DBI extension}
In the soft wall model \cite{Karch:2006pv}, a background metric and dilaton are postulated upon which fluctuations (two gauge fields and a bifundamental scalar) propagate. The backgrounds are given by
\begin{align}
\label{ads}
ds^2 &= \frac{L^2}{z^2}\left(-dt^2 + d\mathbf{x}^2 + dz^2\right)\,,\qquad e^{-\Phi} = e^{-c z^2},
\end{align}
for the low temperature confined phase\footnote{The $z$-coordinate ranges from 0 (the boundary) to $\infty$ (the center of AdS).} and by
\begin{align}
ds^2 &= \frac{L^2}{z^2}\left(-f(z)dt^2 + d\mathbf{x}^2 + \frac{dz^2}{f(z)}\right) \,,\qquad
e^{-\Phi} = e^{-c z^2}
\end{align}
in the high temperature deconfined phase where $f(z) = 1 - z^4/z_h^4$. The radial $z$ coordinate ranges from 0 to $z_h$, the horizon location of the AdS black hole. The temperature of the dual boundary theory is given by the Hawking temperature of the black hole as $T = \frac{1}{\pi z_h}$. \\

Besides the bifundamental scalar (which will be irrelevant for our discussions), the soft wall action consists of two gauge fields with gauge groups $SU(N_f)_L \times SU(N_f)_R$:
\begin{equation}
\label{softwall}
S = - \frac{1}{4g_5^2}\int d^{5}x \sqrt{-g} e^{-\Phi} \text{tr}\left[F^{L,\mu\nu}F_{L,\mu\nu} + F^{R,\mu\nu}F_{R,\mu\nu}\right],
\end{equation}
whose equations of motion are Yang-Mills equations in a curved background (plus dilaton). The original motivation of the soft wall model was to provide an explicit holographic model capable of reproducing the (linear) Regge spectrum of mesons: $m_n^2 \sim \sigma n$, a property which is not that simple to obtain in holographic models (most models obtain $m_n^2 \sim n^2$ (e.g. the hard wall model)). Instead the soft wall model in the AdS background (\ref{ads}) yields $m_n^2 = 4cn$. The price one pays for this is that the soft wall model does not solve Einstein's equations. Since then, models have been constructed that satisfy both of these properties \cite{dePaula:2008fp,Gursoy:2008za,Gursoy:2009jd,Li:2012ay,He:2013qq}, though one pays for this in simplicity. It was the original hope of the authors of \cite{Karch:2006pv} that the model nevertheless would lead to interesting phenomenology and it is with this state of mind that we will proceed. \\

To study the heavy quarkonia, the authors of \cite{Fujita:2009wc,Fujita:2009ca} suggest choosing a flavor-dependent soft-wall parameter $c$, where the light quarks ($u$, $d$, $s$) are combined into a $SU(3)_L\times SU(3)_R$ soft wall model and the heavy quark of interest (charm) is treated on its own in a $U(1)_L \times U(1)_R$ Abelian model:
\begin{equation}\label{fuj}
S = - \int d^5x \sqrt{-g}~\text{tr}~e^{-c_{\rho}z^2}\mathcal{L}_{light} + e^{-c_{J/\psi}z^2}\mathcal{L}_{charm}.
\end{equation}
This proposal immediately dismisses the influence of the heavy-light meson sector (by choosing $\mathfrak{u}(4) \to \mathfrak{u}(3) \oplus \mathfrak{u}(1)$), but as was illustrated in \cite{Fujita:2009wc,Fujita:2009ca}, some suggestive qualitative results can be obtained for the $c\bar{c}$ meson. \\
The light quark soft wall model is fixed by choosing $c_\rho = 0.151$ GeV$^2$. For the heavy charmonium soft wall model, we instead choose $c_{J/\psi} = 2.43$ GeV$^2$. Both of these values have been chosen to fix the mass of the meson to its experimental value: $m_\rho = 0.77$ GeV and $m_{J/\psi} = 3.1$ GeV.\footnote{Actually the values are computed using a slightly shifted mass as is discussed in \cite{Fujita:2009wc,Fujita:2009ca}.} The Lagrangian of the charm sector hence exhibits a stronger damping and it was argued in \cite{Fujita:2009wc,Fujita:2009ca} that it can be neglected to determine the deconfinement temperature, which equals $T_c = 0.492\sqrt{c_{\rho}} = 0.191$ GeV \cite{Herzog:2006ra}. From here on, we hence focus on the $U(1)_L \times U(1)_R$ soft wall model for the charm sector. \\
The idea of \cite{Fujita:2009wc,Fujita:2009ca} is to consider the spectral peaks in the spectral function and to find out, as the temperature is increased, when the bound states disappear (= melt). If this happens above $T_c$, the meson melts only at a higher temperature than the deconfinement temperature. If this happens below $T_c$, the state melts precisely at $T_c$ itself: the reason for this is that for $T<T_c$, the black hole background should not be used anymore since the first-order Hawking-Page transition will have occurred as the temperature is lowered below $T_c$.

In \cite{Fujita:2009wc,Fujita:2009ca}, where the action \eqref{fuj} was originally proposed, a few words were spent on the origin of the 2 scales, $c_\rho$ and $c_{J/\psi}$. A priori, only one scale $c$ should be present since this $c$ is intimately related to the unique principal QCD confinement scale. In principle, the latter should correspond to the string tension, as mentioned in \cite{Fujita:2009wc,Fujita:2009ca}, but the underlying soft wall model does not even support a string tension, in the sense that the Wilson loop does not display an area law to begin with \cite{Karch:2010eg}. The charmonium mass is mostly dictated by the constituent charm quark mass, which is an order of magnitude larger than the confinement scale, nonetheless we fixed, following \cite{Fujita:2009wc,Fujita:2009ca}, the second scale $c_{J/\psi}$ to faithfully reproduce the mass of the $J/\psi$. Analogous to \cite{Fujita:2009wc,Fujita:2009ca}, we consider \eqref{fuj} to constitute a pragmatic first trial to study nontrivial charmonia properties. Similar approaches as in \cite{Fujita:2009wc,Fujita:2009ca} can be found in \cite{Kim:2007rt,Hou:2007uk}.\\
Nonetheless, in soft wall models it can be motivated that $c_{J/\psi}$ will dynamically change its value w.r.t.~$c_\rho$ due to back reaction effects of the heavy quark sector on the geometry. A back reaction study in a hard wall setting \cite{Shock:2006gt} revealed in particular that an effective soft wall model is recovered with a $c$ depending on the heavy quark mass. Quite likely, a similar phenomenon will take place starting with a soft wall model, if we were smart enough to carry out the involved analysis.\\
As was noticed in \cite{Grigoryan:2010pj}, many other papers following the so-called top-down approach to QCD ---where holographic models directly rooted in string theory are used to understand essential QCD features \cite{Kruczenski:2003uq,Hong:2003jm,Mateos:2006nu,Liu:2006nn}--- suffer from the same problem, since only a single scale is governing the dynamics, whereas dual descriptions of charmonia need at least two.\\

In the meantime, more involved holographic charmonium models became available on the market, see e.g.~\cite{Grigoryan:2010pj,Hohler:2013vca}, where additional parameters are introduced to also provide a realistic charmonium decay width etc. This goal is essentially achieved by making a wise choice for the dilaton profile. For now, we leave these for further study, and focus here on the generalization to the magnetic field case of the pioneering model \eqref{fuj} \cite{Fujita:2009wc,Fujita:2009ca}.\\

The charm vector modes are defined via $2V = A_L + A_R$ and these are the ones we are interested in. For our purposes, this action is not sufficient to handle a background magnetic field. Indeed, the linearity of Maxwell's equations implies for the decomposition $F = \bar{F} + \tilde{F}$:
\begin{equation}
\partial_{\mu}\left(e^{-\Phi}\sqrt{-g}F^{\mu\nu}\right) = \underbrace{\partial_{\mu}\left(e^{-\Phi}\sqrt{-g}\bar{F}^{\mu\nu}\right)}_{=0} + \partial_{\mu}\left(e^{-\Phi}\sqrt{-g}\tilde{F}^{\mu\nu}\right) = 0,
\end{equation}
where the first term vanishes due to the fact that a constant magnetic field solves the soft wall Maxwell equations of motion. This means the fluctuations do not feel that the magnetic field is turned on. The $J/\psi$ particle is a charm-anticharm bound state and hence is uncharged. The fact that it does not couple to a magnetic field is quite natural: only the interior quarks feel the presence of the magnetic field.\footnote{In purely classical terms, the $J/\psi$ state is like an electric dipole whose charges are bound to each other by the strong force.} To overcome this feature, one might be tempted to take a non-Abelian gauge field instead (for instance mixing the charm with the top quarks). This would lead to non-Abelian interaction terms present in the equations of motion that can couple a background field to a fluctuation. This is not the right thing to do: the uncharged mesons (the $J/\psi$ and the $\Upsilon$) would still not feel the magnetic field, whereas all we would have accomplished is the study of the influence of the magnetic field on the \emph{charged} mesons ($c\bar{t}$ and $\bar{c}t$ in our example). \\

Instead of doing this, we need to be able to probe the internal structure of the meson with a magnetic field. In the past, the Born-Infeld non-linear electrodynamic model was proposed to overcome the infinite self-energy of a static point-charge \cite{BI} (see also the book \cite{Zwiebach:2004tj} for a nice discussion on this). The crucial feature that emerges is the appearance of a new length scale: point-charges in the theory acquire a self-energy that is of the form of an extended object on the order of this length scale. In string theory these equations reappear where the new length scale is the string length $\ell_s$. Hence in that case, Maxwell's equations hold for scales larger than the string scale, smaller length scales require utilizing the Born-Infeld action (or its D-brane cousins). Since the general spirit of holographic models is to use strings as a guide (these being roughly identified with the gauge strings), we are led to using the BI-action as a plausible guess of an extension of the soft wall model, capable of probing the internal structure of mesons.\\
The DBI action naturally appears in top-down constructions of confining gauge theories, as it describes the fluctuations of the underlying D-branes. Let us for example refer to the Sakai-Sugimoto model (D4/D8) \cite{Sakai:2004cn,Sakai:2005yt} or D4/D6 models of \cite{Kruczenski:2003be,Kruczenski:2003uq}. The mesons of the dual gauge theory are identified with these fluctuations. Also in these setups, submesonic dynamics can be taken into account. To illustrate this, let us mention that in \cite{Callebaut:2011ab,Callebaut:2013wba}, it was found that a magnetic field has an influence on the vector ($\rho$) meson spectrum, this partially due to the quark constituents' response to the magnetic field. In different work, the internal meson structure was studied via the form factors/parton distribution functions \cite{Vega:2009zb,Vega:2010ns,BallonBayona:2009ar,Koile:2011aa,Koile:2013hba,Koile:2014vca}, see also the review \cite{Brodsky:2014yha}. Such analyses can be carried out by either using bottom-up hard/soft wall models, the top-down models mentioned before or the Polchinski-Strassler background \cite{Polchinski:2000uf,Polchinski:2001tt}. \\

The action we use is the DBI-version of the soft wall model
\begin{equation}
S = -\frac{1}{4\pi^2\alpha'g_5^2}\int d^{D}x e^{-\Phi}\sqrt{-\det\left(g_{\mu\nu}+2\pi\alpha' i F_{L,\mu\nu}\right)} + (F_L \leftrightarrow F_R)
\end{equation}
for left- and right-handed $U(1)$ gauge bosons. The prefactors have been chosen in such a way as to precisely yield the correct prefactor for the $F^2$ term in the expansion. Thus this action can be seen as a generalization of the conventional soft-wall model (\ref{softwall}) in which interactions of the gauge field have been included in this specific form. We take the gauge field to be a background plus a fluctuation $F = \bar{F} + \tilde{F}$ and we determine the equations of motion for non-self-interacting fluctuations (quadratic in the action\footnote{At this level, vector and axial vector modes do not mix due to their different intrinsic parity. In the same context, it is also useful to point out that the pseudoscalar charmonium mode, the $\eta_c$, could also mix with the $J/\psi$, see \cite{Machado:2013rta,Alford:2013jva}, but again due to the intrinsic parity, this can at the (holographic) QCD-level only happen due to the Chern-Simons piece of the action (which represents the triangle anomaly). In the current work we will ignore this Chern-Simons contribution as it is relatively suppressed w.r.t.~the DBI-action with an extra (supposedly large) 't Hooft coupling constant. For a similar comment at the level of $\rho$ and $\pi$ mesons, let us refer to \cite{Callebaut:2011ab,Callebaut:2013wba}. In the context of a Bethe-Salpeter equation analysis, an anomalous mixing between $\eta_c$ and $J/\psi$ by means of a magnetic field was also reported in \cite{Cho:2014loa}. }), but exact in the background field. In writing down this action, we have implicitly neglected all forms of interaction between left and right modes. We restrict our study to only vector-modes, hence both actions above are actually the same.\footnote{When rewriting the above action in terms of vector and axial modes, explicit coupling terms are present. However, due to the symmetry $L \leftrightarrow R$, terms in the Lagrangian only have an even number of axial gauge fields. Since we are restricting ourselves to the part quadratic in the fluctuations, the axial part has no influence at all on the vector part we are interested in. Note that when including the background magnetic field in the vector part, terms with field factors such as $V^m A$ might be worrisome, even for the quadratic part. Luckily such terms are absent.} \\

It is important to emphasize that, for vanishing background electromagnetic field, the (quadratic part of the) model reduces to the normal soft-wall Maxwell action. Thus it encompasses the successes of the latter but still allows interactions with a background field in the bulk which are controlled by a further (dimensionful) parameter $\alpha'$.

\section{Setup of the model}
Before delving into the computations of the fluctuations in this model, let us first set up the model in a proper way. We will show that a constant magnetic field is in fact a solution to the soft-wall DBI equations of motion (without backreaction). After that, we will present a method to fix the additional parameter $\alpha'$.

\subsection{A constant magnetic field solves the DBI equations of motion in the soft wall background}
A constant background magnetic field can be modeled by turning on a constant $F_{12} = - F_{21}$ while keeping the other components zero. Our first goal is to demonstrate that this background magnetic field solves the DBI equations of motion (EOMs) in the soft wall background. We start with
\begin{equation}
S \sim \int d^{D}x e^{-\Phi}\sqrt{-\det(g+2\pi\alpha' iF)}
\end{equation}
for a fixed background dilaton and (string) metric. Varying this action with respect to $A_{\mu}$ yields
\begin{equation}
\delta S \sim -\int d^{D}x \left(\frac{1}{g+2\pi\alpha' iF}\right)^{\nu\lambda}\delta F_{\lambda\nu}e^{-\Phi}\sqrt{-\det(g+2\pi\alpha' iF)},
\end{equation}
which gives
\begin{equation}
\partial_{\lambda}\left[\left(\frac{1}{g+2\pi\alpha' iF}\right)^{\nu\lambda}e^{-\Phi}\sqrt{\det(g+2\pi\alpha' iF)} - \left(\frac{1}{g+2\pi\alpha' iF}\right)^{\lambda\nu}e^{-\Phi}\sqrt{\det(g+2\pi\alpha' iF)}\right] = 0.
\end{equation}
The specific form of $g$, $\Phi$ and the candidate $F$ (only 1,2 and 2,1 indices which are constant), shows that nothing is dependent on any other coordinate than $z$. Thus $\lambda = z$ is the only possibility. But if $\lambda = z$, $\nu$ should be $z$ as well (due to the specific form of $\frac{1}{g+2\pi\alpha' iF}$). But then both terms cancel each other and indeed this candidate $F$ solves the EOMs in this particular soft wall background.

\subsection{Fixing the string length parameter $\alpha'$}
\begin{figure}[h]
\centering
\includegraphics[width=0.35\linewidth]{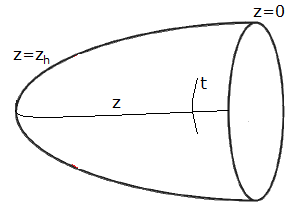}
\caption{Shape of the surface spanning a thermal loop in the boundary theory.}
\label{poly}
\end{figure}

Let us next try to fix the additional parameter $\alpha'$ that we introduced in the previous section. It is known \cite{Karch:2010eg} that the soft-wall model background does not predict a satisfactory Wilson loop (which is the prototypical way to fix $\alpha'$). Another quantity that might allow us to fix $\alpha'$ is the Polyakov loop. Its value has been computed holographically for several theories \cite{Andreev:2006ct,Andreev:2009zk} and shows remarkable similarity to lattice results. We compute the vacuum expectation value of the Polyakov loop in terms of a surface spanning a thermal loop at the boundary. To this effect, we evaluate the on-shell Nambu-Goto action for a string surface wrapping the thermal boundary circle. The resulting surface is cigar-shaped and depicted in FIG.~\ref{poly}.

\noindent The Euclidean Nambu-Goto action,\footnote{The metric to be used here is the string metric, \emph{not} the Einstein metric.}
\begin{equation}
S = \frac{1}{2\pi \alpha'}\int d^{2}\sigma \sqrt{\det{G_{\mu\nu}\partial_{\alpha}X^{\mu}\partial_{\beta}X^{\nu}}},
\end{equation}
can be evaluated in worldsheet coordinates aligned with the space coordinates as $\sigma_1 = \tau$ and $\sigma_2 = z$. We look for static solutions. For the AdS$_5$ black hole metric, this gives
\begin{equation}
\label{act}
S = \frac{L^2}{2\pi \alpha' T}\int_{0}^{z_{h}} d z \frac{1}{z^2}\sqrt{1 + f \mathbf{x}'^2},
\end{equation}
where we expressed the surface as functions $\mathbf{x}(z)$. The extremal configuration hence satisfies
\begin{equation}
\left(\frac{f\mathbf{x}'}{\sqrt{1 + f \mathbf{x}'^2}}\right)' = 0,
\end{equation}
or $\mathbf{x} = \mathbf{C}$, a constant vector.\footnote{This could be observed more quickly by noting that (\ref{act}) has an absolute minimum when considering configurations for which $\mathbf{x} = \mathbf{C}$.} The on-shell action hence becomes
\begin{equation}
S_{\text{on-shell}} = \frac{L^2}{2\pi \alpha' T}\int_{0}^{z_{h}} d z \frac{1}{z^2}.
\end{equation}
After introducing a cut-off at $z=\epsilon$, the Polyakov loop becomes
\begin{equation}
\left\langle L(T)\right\rangle = e^{-S} = e^{\frac{L^2}{2\alpha'} - \frac{L^2}{2\pi \alpha' T \epsilon}}.
\end{equation}
In general, the resulting expression in the exponential consists of two parts: one part is proportional to the circumference of the loop in the boundary theory whereas the other is not. The part that scales with the circumference at the boundary is multiplied by a divergent quantity and the prescription instructs us to drop this part \cite{Maldacena:1998im}, see also \cite{Dumitru:2003hp} for a renormalization analysis of the Polyakov loop. We hence renormalize this expression by subtracting the $1/\epsilon$ divergence so that we obtain
\begin{equation}
\left\langle L(T)\right\rangle_{\text{ren}} = e^{\frac{L^2}{2\alpha'}}.
\end{equation}
Our holographic exercise shows that in this model, the Polyakov loop is temperature independent (which is of course not exactly what is seen from lattice studies, the most accurate way to effectively compute the Polyakov loop). Our best bet is then to take the asymptotically constant value and fit our result to this. Lattice computations \cite[FIG.~4]{Kaczmarek:2002mc} do indeed show that for large $T$ ($T>3T_c$), $\left\langle L(T)\right\rangle$ is about $1.1$ . Using this value immediately yields
\begin{equation}
\frac{\ell_s}{L} \approx 2.29.
\end{equation}
This is sufficient since, knowing that the metric is proportional to $L^2$, the DBI action can be rewritten as
\begin{equation}
\sqrt{-\det\left(g_{\mu\nu} + 2\pi (2.29)^2 L^2 i F_{\mu\nu}\right)} = L^{5} \sqrt{-\det\left(\tilde{g}_{\mu\nu} + 2\pi (2.29)^2 i F_{\mu\nu}\right)}
\end{equation}
and all powers of the AdS length drop out from the equations of motion. We can hence simply set it to $L=1$ (like for the normal soft wall model). The important feature (for which no computation is necessary) is that the string length and the AdS length are of the same order: the (inverse) QCD scale. \\\noindent It is perhaps useful to remind the reader here that the Poyakov loop, just as the Wilson loop, is typically computed for a heavy quark pair, this to avoid of having to deal with the issue of dynamical string breaking. It is from this perspective thus rather natural to fix the new scale we introduced, $\alpha'$, using the (heavy) charm DBI-action.

\section{Equations of motion of fluctuations}
In general one can expand
\begin{align}
&\sqrt{-\det\left(g_{\mu\nu}+2\pi\alpha' i F_{\mu\nu}\right)} = \sqrt{-\det\left(g_{\mu\nu}+2\pi\alpha' i (\bar{F}_{\mu\nu}+F_{\mu\nu})\right)} = \sqrt{-\det\left(\bar{a}_{\mu\nu}+2\pi\alpha' iF_{\mu\nu}\right)} \nonumber\\
&\approx \sqrt{-\det(\bar{a})}\left\{1 + \frac{2\pi\alpha' i}{2}\text{Tr}(\bar{a}^{-1}F) + \frac{(2\pi\alpha' i)^2}{8}(\text{Tr}(\bar{a}^{-1}F))^2 - \frac{(2\pi\alpha' i)^2}{4}\text{Tr}((\bar{a}^{-1}F)^2) + \hdots\right\}.
\end{align}
The first term (after the 1) vanishes by the equations of motion of the background $\mathbf{B}$-field. The other terms are quadratic in the fluctuations. The inverse background $\bar{a}^{-1}$ can be split in a symmetric and antisymmetric part: $S+J$. \\

\noindent The symmetric part only has a contribution from the final term and this is the Maxwell action where indices are raised and lowered with only the symmetric part $S$. \\
\noindent The antisymmetric part looks as follows (where only the $1,2$ and $2,1$ indices of $J$ are present):
\begin{align}
\frac{1}{8}J^{\mu\nu}F_{\nu\mu}J^{\rho\sigma}F_{\sigma\rho} - \frac{1}{4}J^{\mu\nu}F_{\nu\rho}J^{\rho\sigma}F_{\sigma\mu} = 0.
\end{align}
Hence only the symmetric part of the inverse background contributes to the raising and lowering of indices.\footnote{Some simple index gymnastics shows that the term $\text{tr}(SFJF)$ vanishes identically.}
The equations of motion are
\begin{equation}
\label{master}
\partial_{\mu}\left(e^{-\Phi}\sqrt{-\mathcal{G}}F^{\mu\nu}\right)=0,
\end{equation}
where $\mathcal{G}_{\mu\nu} = \bar{a}_{\mu\nu}$. \\

\noindent The coordinates are denoted as $t, x_1, x_2, x_3$ for the dual coordinates and $z$ for the holographic coordinate. More concretely, we take
\begin{equation}
\bar{F}_{12} = - \bar{F}_{21} = \partial_1 A_2 = -iqB\frac{2}{3}
\end{equation}
since the charm quark charge is $+2/3 q$. We get
\begin{eqnarray}
\mathcal{G}_{\mu\nu}=\left[\begin{array}{ccccc}
g_{00} & 0 & 0 & 0 & 0\\
0 & g_{11} & 2\pi\alpha' i \bar{F}_{12} & 0 & 0  \\
0 & -2\pi\alpha' i \bar{F}_{12} & g_{22} & 0 & 0\\
0 & 0 & 0 & g_{33} & 0 \\
0 & 0 & 0 & 0 & g_{zz} \end{array}\right]
\end{eqnarray}
with its determinant
\begin{equation}
\mathcal{G} = g_{00}g_{33}g_{zz}\left(g_{11}g_{22} - (2\pi\alpha')^2\bar{F}_{12}^{2}\right)
\end{equation}
and inverse
\begin{eqnarray}
\mathcal{G}^{\mu\nu}=\left[\begin{array}{ccccc}
\frac{1}{g_{00}} & 0 & 0 & 0 & 0\\
0 & \frac{g_{22}}{X} & - \frac{2\pi\alpha' i \bar{F}_{12}}{X} & 0 & 0  \\
0 & \frac{2\pi\alpha' i \bar{F}_{12}}{X} & \frac{g_{11}}{X} & 0 & 0\\
0 & 0 & 0 & \frac{1}{g_{33}} & 0 \\
0 & 0 & 0 & 0 & \frac{1}{g_{zz}} \end{array}\right]
\end{eqnarray}
where $X = g_{11}g_{22} - (2\pi\alpha')^2\bar{F}_{12}^{2}$. \\

\noindent We choose the gauge $A_{z} = 0$ and focus on vector modes $V = \frac{A_L + A_R}{2}$.\footnote{Note that we refrain from making an additional gauge choice such as $\partial^{\mu} A_{\mu} = 0$ as was done by \cite{Karch:2006pv,Fujita:2009wc,Fujita:2009ca} because there appear to be some subtleties in this choice for the AdS black hole. For the polarizations and momentum choice we will make shortly, we actually do not need to impose an additional gauge choice to simplify the equations of motion. We will come back to this issue in a forthcoming paper.}
We will denote by $G$ only the symmetric part of the metric tensor $\mathcal{G}$:
\begin{eqnarray}
G^{\mu\nu}=\left[\begin{array}{ccccc}
\frac{1}{g_{00}} & 0 & 0 & 0 & 0\\
0 & \frac{g_{22}}{X} & 0 & 0 & 0  \\
0 & 0 & \frac{g_{11}}{X} & 0 & 0\\
0 & 0 & 0 & \frac{1}{g_{33}} & 0 \\
0 & 0 & 0 & 0 & \frac{1}{g_{zz}} \end{array}\right].
\end{eqnarray}
The equations of motion follow from (\ref{master}).
Next, we make a Fourier expansion of the modes $\sim e^{i\mathbf{q}\cdot \mathbf{x} - i \omega t}$. In the remainder of this paper, we will focus on the case $\mathbf{q}=0$ (no momentum on the boundary). We hope to return to the momentum-dependent part in the future. \\

\noindent The equation of motion for $V_1$ is given by
\begin{equation}
\label{V1par}
\partial^2_{z}V_{1} + \partial_z\left(\ln\left(\sqrt{-\mathcal{G}}e^{-cz^2}G^{zz}G^{11}\right)\right)\partial_z V_1 - \frac{G^{tt}}{G^{zz}}\omega^2 V_1 = 0
\end{equation}
(and exactly the same equation of motion for $V_2$). \\
$V_3$ obeys
\begin{equation}
\label{V3par}
\partial^2_{z}V_{3} + \partial_z\left(\ln\left(\sqrt{-\mathcal{G}}e^{-cz^2}G^{zz}G^{33}\right)\right)\partial_z V_3 - \frac{G^{tt}}{G^{zz}}\omega^2 V_3 = 0
\end{equation}
which is different since $G^{33}$ is of a simpler form.

\section{Numerical solution of the spectral function}
\subsection{Asymptotics: Frobenius analysis}
The numerical recipe we will use follows closely the work of \cite{Teaney:2006nc,Fujita:2009wc,Fujita:2009ca}. To start the numerical procedure, we need the asymptotics of the solutions. We first look into the horizon region and then into the boundary region.

\subsubsection{Near-horizon limit}
\noindent We start by looking at fluctuations whose polarization is parallel to the applied magnetic field (\ref{V3par}). Let us denote $D = 2\pi \alpha' i \bar{F}_{12}$. Then
\begin{align}
\ln\left(\sqrt{-\mathcal{G}}e^{-cz^2}G^{zz}G^{33}\right) &= \ln\left(\sqrt{\frac{L^{10}}{z^{10}}+\frac{L^6}{z^6}D^2}\frac{z^4}{L^4}\left(1-\frac{z^4}{z_h^4}\right)\right) - cz^2, \\
\frac{G^{tt}}{G^{zz}} &= -1/f^2.
\end{align}
To proceed further, it is convenient to rescale the differential equation to eliminate $c$ from the equation by setting:
\begin{equation}
\xi = \sqrt{c}z, \quad \tilde{\omega} = \frac{\omega}{\sqrt{c}}, \quad \tilde{D} = \frac{D}{c}.
\end{equation}
We can define a dimensionless temperature associated to these parameters as $t = \frac{T}{\sqrt{c}}$. The deconfinement temperature then becomes $t_c = \frac{T_c}{\sqrt{c_{J/\psi}}} = 0.122$. As discussed previously, for $t$ lower than this value, one should in fact resort to the thermal AdS background instead. Just like the authors of \cite{Fujita:2009wc,Fujita:2009ca}, we will still show results with $t < t_c$, in part because the identification of the spectral peaks will be more clear. The conclusions we will obtain in the end will be independent of this choice. Other reasons to look into this were discussed in \cite{Fujita:2009wc,Fujita:2009ca}: the deconfinement temperature may be smaller than $t_c$ since the soft-wall model does not solve Einstein's equations. Also the phase transition in real QCD is expected not to be sharp and the results for the black hole background at $t<t_c$ may provide indications of this smooth behavior. \\
The above rescaling transforms (\ref{V3par}) into
\begin{equation}
\label{V3parnew}
\partial^2_{\xi}V_{3} + \partial_\xi\left(\ln\left(\sqrt{\frac{L^{10}}{\xi^{10}}+\frac{L^6}{\xi^6}\tilde{D}^2}\frac{\xi^4}{L^4}\left(1-\frac{\xi^4}{\xi_h^4}\right)\right)-\xi^2 \right)\partial_\xi V_3 +\left(\frac{\tilde{\omega}^2}{f^2}\right)V_3 = 0.
\end{equation}

\noindent The only term worth discussing is the second one (containing a single $\partial_\xi$). In the limit $\xi \to \xi_h$, the prefactor of this term becomes
\begin{equation}
\frac{-4\xi_h^4}{\xi_h(\xi_h^4-\xi^4)}
\end{equation}
and this is precisely of the same form as when $\mathbf{B}=\mathbf{0}$: the background magnetic field does not influence the near-horizon behavior of this term in the differential equation.
Substituting $V_3 = \left(1-\frac{\xi}{\xi_h}\right)^\alpha$ in the fluctuation equation (\ref{V3parnew}), we hence obtain the same indicial equation as in the case $\mathbf{B}=\mathbf{0}$ leading to
\begin{equation}
\alpha = \pm i \frac{\tilde{\omega} \xi_h}{4}.
\end{equation}

\noindent Quite analogously, when the polarization is perpendicular to the applied magnetic field, the (rescaled) governing equation is given by
\begin{align}
\label{V1parnew}
\partial^2_{\xi}V_{1} &+ \partial_\xi\left(\ln\left(\frac{\sqrt{\frac{L^{10}}{\xi^{10}}+\frac{L^6}{\xi^6}\tilde{D}^2}}{\frac{L^4}{\xi^4}+\tilde{D}^2}\left(1-\frac{\xi^4}{\xi_h^4}\right)\right)-\xi^2 \right)\partial_\xi V_1 +\left(\frac{\tilde{\omega}^2}{f^2} \right)V_1 = 0,
\end{align}
where the complicated prefactor in the second term, in the limit $\xi \to \xi_h$, becomes analogously
\begin{equation}
-\frac{4\xi_h^4}{\xi_h(\xi_h^4-\xi^4)}
\end{equation}
which is again of the same form as in the case with $\mathbf{B}=\mathbf{0}$. \\

\noindent To summarize, a Frobenius analysis around the horizon ($\xi \to \xi_h$) shows that the asymptotics for both of these differential equations is given by
\begin{equation}
V \sim \left(1-\frac{\xi}{\xi_h}\right)^{\pm i \tilde{\omega} \frac{\xi_h}{4}},
\end{equation}
which is hence not altered by turning on a background magnetic field. Note that since $\tilde{D}^2>0$, no extra singularity is created at all by turning on a magnetic field.

\subsubsection{Near-boundary limit}
\noindent In the boundary limit $\xi\to 0$, we take $\xi^\alpha$ as a first term in the power series analysis and we find that the second terms in both (\ref{V3parnew}) and (\ref{V1parnew}) behave as $-1/\xi$, which is the same as in the case of $\mathbf{B}=\mathbf{0}$ (every $\tilde{D}$ multiplies at least one $\xi$). $\frac{G^{33}}{G^{zz}}$ and $\frac{G^{11}}{G^{zz}}$ behave as $1$ (and therefore cannot create more divergent terms).
The asymptotic analysis then simply gives
\begin{equation}
\alpha(\alpha-1) - \alpha = 0,
\end{equation}
yielding $\alpha = 0$ or $\alpha = 2$. In this case, we require more (since we will solve the differential equations by evolving from the boundary to the horizon): we need the first derivatives of $\Phi$ at the holographic boundary. For this, the Frobenius analysis needs to be taken one level further. The indicial equation has two solutions whose difference is an integer. The largest of the two (= 2) has a regular series expansion, whereas the other solution has a logarithmic series as well:
\begin{align}
\Phi_2(\xi) &= \xi^2\sum_{k=0}^{+\infty}a_k \xi^k, \nonumber\\
\Phi_1(\xi) &= C\ln(\xi)\Phi_2(\xi) + \sum_{k=0}^{+\infty}b_k \xi^k.
\end{align}
As normalization, we choose $a_0=b_0=1$. Also $b_2$ is a free parameter, since it corresponds to simply adding $\Phi_2$ to $\Phi_1$ with factor $b_2$. Explicit analysis in our case (for both cases) shows that $a_2 = -\frac{\tilde{\omega}^2}{8}+\frac{1}{2}$ and $C = - \frac{\tilde{\omega}^2}{2}$.\\
All odd-indexed parameters are automatically zero.
The crucial information is then encoded in
\begin{align}
\Phi_2'(\epsilon) &= 2 \epsilon , \nonumber\\
\Phi_1'(\epsilon) &= \left.\partial_\xi \left(C\ln(\xi)\xi^2 + F\xi^2\right)\right|_{\xi=\epsilon},
\end{align}
where we can freely choose $F$ as we will.

\subsection{Real-time holographic prescription}
The real-time holographic prescription \cite{Son:2002sd,Policastro:2002se} requires us to construct the ingoing solution at the horizon. In the above, we have constructed two linearly independent real solutions $\Phi_1$ and $\Phi_2$ with the boundary conditions:\footnote{In \cite{Fujita:2009wc,Fujita:2009ca}, the boundary conditions are instead
\begin{align}
\Phi_1(\epsilon) &= 1 , \quad \Phi_1'(\epsilon) = -\tilde{\omega}^2\ln\left(\tilde{\omega}\frac{\epsilon}{2}\right)\epsilon - \epsilon \gamma \tilde{\omega}^2, \nonumber \\
\Phi_2(\epsilon) &= \epsilon^2, \quad \Phi_2'(\epsilon) = 2\epsilon,
\end{align}
with $\gamma$ the Euler-Mascheroni constant. The extra terms added to $\Phi_1'(\epsilon)$ (corresponding to choosing $F$ in the formulas above) are chosen such that the asymptotic Bessel functions are found. This is not necessary and we believe things are more transparent by simply dropping these terms.
}
\begin{align}
\Phi_1(\epsilon) &= 1 , \quad \Phi_1'(\epsilon) = -\tilde{\omega}^2 \ln\left(\epsilon\right)\epsilon, \nonumber\\
\Phi_2(\epsilon) &= \epsilon^2, \quad \Phi_2'(\epsilon) = 2\epsilon.
\end{align}
Since also the ingoing and outgoing solutions (with behavior $\phi_{\pm} \sim \left(1-\frac{\xi}{\xi_h}\right)^{\pm i \tilde{\omega} \frac{\xi_h}{4}}$ near the horizon) are a complete set, we can set
\begin{align}
\Phi_1 &= \alpha \phi_- + \alpha^* \phi_+ \nonumber\\
\Phi_2 &= \beta \phi_- + \beta^* \phi_+.
\end{align}
If we normalize the desired solution $v$ as
\begin{equation}
v = \Phi_1 + B \Phi_2 = (\alpha+ B\beta)\phi_- + \underbrace{(\alpha^* + B\beta^*)}_{0}\phi_+,
\end{equation}
we obtain $B = -\frac{\alpha^*}{\beta^*}$.

\noindent The quadratic fluctuations come from a Maxwell-type action:
\begin{equation}
S \sim \int d^{5}x e^{-\Phi}\sqrt{-\mathcal{G}}G^{\mu\rho}G^{\nu\sigma}F_{\mu\nu}F_{\rho\sigma}.
\end{equation}
It is immediate that the bulk action vanishes on-shell. The boundary contribution (obtained by integrating by parts and retaining \emph{only} the contribution at $z=0$ (cf.~the real-time prescription of \cite{Son:2002sd,Policastro:2002se})) is schematically given by
\begin{equation}
S_{\text{on-shell, bdy}} \sim \lim_{z \to 0} \int d^{4}x \sqrt{-\mathcal{G}}G^{zz}G^{\nu\sigma}A_{\nu}\partial_z A_{\sigma}.
\end{equation}
We need the $z\approx0$ behavior of several quantities:
\begin{align}
G^{zz} &= \frac{f z^2}{ L^2} \approx \frac{z^2}{L^2} + \mathcal{O}(z^6),\nonumber\\
G^{tt} &= \frac{z^2}{f L^2} \approx \frac{z^2}{L^2} + \mathcal{O}(z^6),\nonumber\\
G^{11} &= \frac{\frac{L^2}{z^2}}{\frac{L^4}{z^4}+\tilde{D}^2} \approx \frac{z^2}{L^2} + \mathcal{O}(z^6),\nonumber \\
G^{33} &= \frac{z^2}{L^2} ,\nonumber\\
\sqrt{-\mathcal{G}} &= \sqrt{\frac{ L^{10}}{z^{10}}+\frac{L^6}{z^6}\tilde{D}^2} \approx \frac{L^5}{z^5} + \mathcal{O}(1/z).
\end{align}
Since the inverse metric $G^{\nu\sigma}$ is diagonal, this action splits into the sum of the different polarizations. Without paying attention to $z$-independent prefactors, we can write
\begin{equation}
S_{\text{on-shell, bdy}} \sim \lim_{z \to 0} \int d^{4}x \frac{L^5}{z^5} \frac{z^2}{L^2} \frac{z^2}{L^2} A_{i}\partial_z A_{i} = \lim_{z \to 0} \int d^{4}x \frac{L}{z} v\partial_z v
\end{equation}
or in terms of $\xi$,\footnote{A global factor of $c$ is needed in this conversion, but we will not keep track of prefactors.}
\begin{equation}
S_{\text{on-shell, bdy}} \sim \lim_{\xi \to 0} \int d^{4}x \frac{L}{\xi} v\partial_\xi v,
\end{equation}
where we only retained the most singular contributions. If this is finite, the subdominant terms will all vanish as $\xi\to 0$. \\

\noindent The prescription instructs us to Fourier transform in the 4d boundary theory and then extract the $\xi$-direction parts. If the field has harmonic dependence $\sim e^{i\mathbf{q}\cdot \mathbf{x} - i \omega t}$ (as is the case here), this simplifies to
\begin{equation}
\lim_{\xi \to 0} \int d^{4}x \frac{L}{\xi} v\partial_z v = \lim_{\xi \to 0} V_4 \frac{L}{\xi} \tilde{v}\partial_\xi \tilde{v},
\end{equation}
where by the tilde-notation we denote that the plane wave part has been extracted. The prescription reduces to simply dropping $V_4$ in this case. Thus
\begin{equation}
D_R (\omega, \mathbf{q}) \sim \lim_{\xi \to 0} \frac{\tilde{v}\partial_\xi \tilde{v}}{\xi}.
\end{equation}

\noindent Up to a proportionality constant, the imaginary part of the retarded Green function is the spectral function and hence
\begin{equation}
\rho(\omega,\mathbf{q}) = -\frac{\Im D_{R}(\omega, q)}{\pi} \sim \Im\left(\frac{(1+B\epsilon^2)(\Phi_1'(\epsilon) + 2B\epsilon)}{\epsilon}\right) = 2\Im B.
\end{equation}
In the final equality we have used the small $\epsilon$ limit. Hence the imaginary part of $B$ holds the information on the spectral function and it is this quantity that we will display in the following paragraphs.

\subsection{Numerical results}

\begin{figure}[h]
\centering
\includegraphics[width=0.7\linewidth]{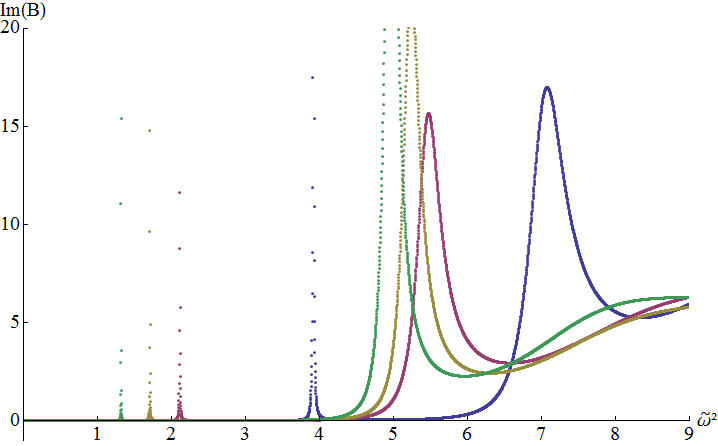}
\caption{Spectral function for the polarization parallel to $\mathbf{B}$ as a function of $\tilde{\omega}^2$ for different values of $qB$. The temperature is fixed at $t=0.07$. Blue: $qB=0$, purple: $qB = 0.2$ GeV$^2$, yellow: $qB = 0.4$ GeV$^2$, green: $qB = 1.0$ GeV$^2$.}
\label{spf1}
\end{figure}

Firstly, we focus on the case when the polarization is parallel to $\mathbf{B}$. The spectral function as a function of $\omega^2$ is shown in FIG.~\ref{spf1} for varying background magnetic field strengths with a fixed temperature $t=0.07$.

\noindent It is clear that spectral peaks are monotonically shifted towards lower $\omega^2$. The critical value $\omega^2=0$ is however never reached for any peak no matter how strong the applied field. This is fortunate since otherwise tachyonic instabilities might set in.\\

\begin{figure}[h]
\centering
\includegraphics[width=0.7\linewidth]{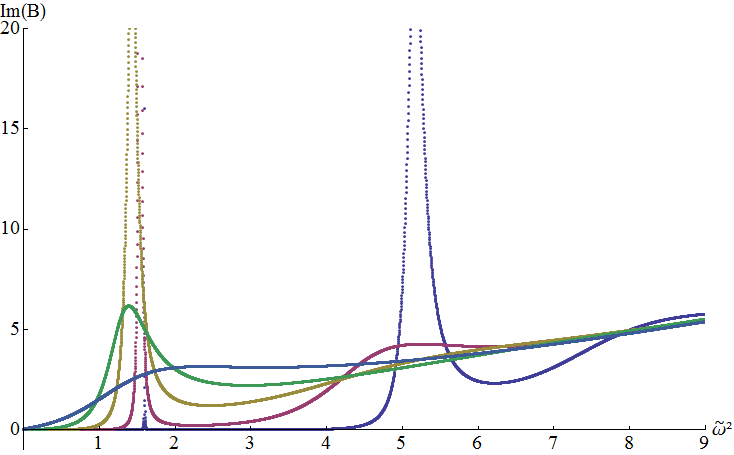}
\caption{Spectral function for the polarization parallel to $\mathbf{B}$ as a function of $\tilde{\omega}^2$ for different values of $t$. The magnetic field is fixed at $qB=0.5$ GeV$^2$. Blue: $t=0.07$, purple: $t = 0.09$, yellow: $t = 0.11$, green: $t = 0.14$, lightblue: $t = 0.20$.}
\label{spf2}
\end{figure}

\noindent In FIG.~\ref{spf2} below we show the same figure but with varying temperature and fixed background magnetic field $qB=0.5$ GeV$^2$.

\noindent From this figure one readily deduces that the spectral peaks only melt at higher temperatures than if $\mathbf{B}=\mathbf{0}$. In the case of FIG.~\ref{spf2}, the lowest lying peak is found to melt only at around $t=0.20$, which is substantially higher than the melting temperature $t=0.14$ when $\mathbf{B}=\mathbf{0}$.

\begin{figure}[h]
\centering
\includegraphics[width=0.7\linewidth]{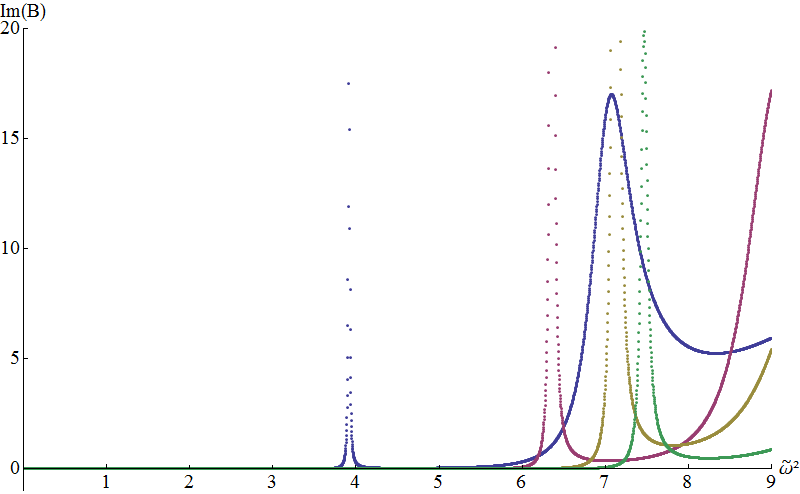}
\caption{Spectral function for the polarization transverse to $\mathbf{B}$ as a function of $\tilde{\omega}^2$ for different values of $qB$. The temperature is fixed at $t=0.07$. Blue: $qB=0$, purple: $qB = 0.2$ GeV$^2$, yellow: $qB = 0.4$ GeV$^2$, green: $qB = 1.0$ GeV$^2$.}
\label{spf3}
\end{figure}

\begin{figure}[h]
\centering
\includegraphics[width=0.7\linewidth]{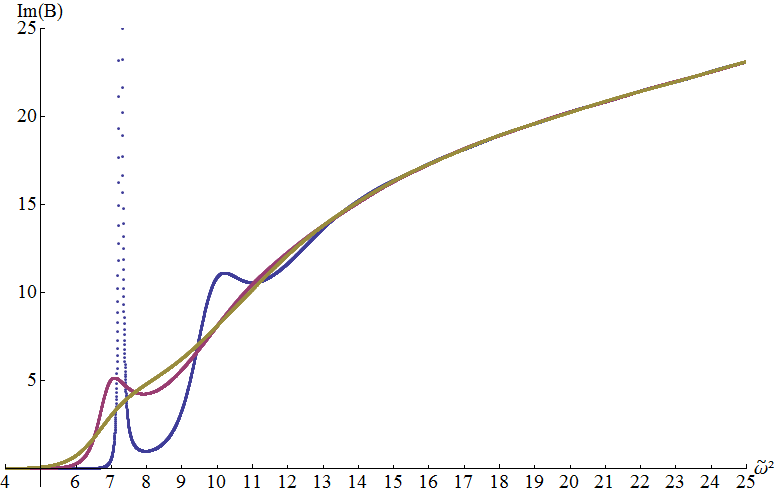}
\caption{Spectral function for the polarization transverse to $\mathbf{B}$ as a function of $\tilde{\omega}^2$ for different values of $t$. The magnetic field is fixed at $qB=0.5$ GeV$^2$. Blue: $t=0.07$, purple: $t = 0.09$, yellow: $t = 0.11$.}
\label{spf4}
\end{figure}

\noindent The spectral function for fluctuations with polarization transverse to the applied magnetic field is shown in FIGS.~\ref{spf3} and \ref{spf4}.

\noindent It is clear that in this case, the behavior is opposite to that of the longitudinal polarizations: peaks shift towards larger $\omega^2$ and they melt at lower temperatures. As the magnetic field increases, the spectral peaks do not shift without bound towards larger $\omega^2$: they approach a fixed asymptotic value that we will discuss further in the next section. It is reassuring to note that a sum rules approach at zero temperature, \cite{Cho:2014loa}, confirms the picture of relatively faster rising masses for the transversal polarizations.

\subsection{Thermal AdS ($f=1$)}

Setting $f=1$, the space reduces to thermal AdS space. In this case, just as with zero magnetic field  ($\mathbf{B} = \mathbf{0}$), we expect to find a discrete spectrum of normalizable modes whose spectral peaks are Dirac-peaks which coincide with the widened peaks determined above. The fluctuation equations reduce to the form:
\begin{align}
\partial^2_\xi v + \left(\frac{(L^4 - \tilde{D}^2\xi^4)}{(-L^4 - \tilde{D}^2\xi^4)\xi} - 2\xi\right) \partial_\xi v  + \tilde{\omega}^2 v &= 0, \quad \text{longitudinal},\\
\partial^2_\xi v + \left(\frac{(L^4 + 3\tilde{D}^2\xi^4)}{(-L^4 - \tilde{D}^2\xi^4)\xi} - 2\xi\right) \partial_\xi v  + \tilde{\omega}^2 v &= 0, \quad \text{transversal}.
\end{align}
\noindent Note that these eigenfunctions should be normalized as dictated by the action itself; we can rescale the eigenfunctions to conventionally normalized modes as $u_i= \sqrt{e^{-\xi^2}\sqrt{-\mathcal{G}}G^{zz}G^{ii}} v_i$. Incidentally, these modes also satisfy a Schr\"odinger equation. \\

\noindent It is found that the discrete spectrum at very high magnetic field asymptotes to
\begin{equation}
\tilde{\omega}^2 = 4n , \quad n=0,1,\hdots
\end{equation}
for the polarization parallel to the magnetic field and
\begin{equation}
\tilde{\omega}^2 = 8 + 4n , \quad n=0,1,\hdots
\end{equation}
for the polarizations transverse to the magnetic field. We believe these to be the asymptotic values of the spectral peaks discussed in the previous section, a feature that can be numerically verified. \\

\noindent So far, in this work, we have focussed on a specific model (DBI) to correct the non-interacting Maxwell equations in the bulk. Of course, there exists an entire array of possibilities here. One other approach might be to insert a $F^4$ correction to the soft wall Maxwell action. How would this alter our conclusions here? \\
\noindent In our formulas, letting $\alpha'$ be small (which for our purposes is exactly the same as choosing a small magnetic field), after Taylor expanding the DBI action, we can make contact with the low $\alpha'$ limit of for instance solely a $F^4$ correction to the action. We have explicitly verified that similar peak-shifting effects are observed (as the shifting of the peaks happens monotonously in $\mathbf{B}$). Hence, we conclude that this effect (at least for small magnetic fields) seems to be quite generic and is not just a peculiarity of this specific extension of the soft wall model.

\section{Comparison to models and the lattice}
In \cite{Chernodub:2010bi}, it was argued that for magnetic fields perpendicular to the gauge string, the string becomes more and more unlikely to break, effectively increasing the string tension. On the lattice \cite{Bonati:2014ksa} it was observed that, when performing this experiment (moving a $Q\bar{Q}$ pair apart from each other), when the axis connecting both quarks is parallel to the applied magnetic field, the string tension decreases and the opposite happens for a perpendicular orientation.

It is not that easy to link this result for the static $Q\bar{Q}$ pair to our result in which we only consider the effective mesonic degrees of freedom and not the individual quarks (whose influence we hoped to have modeled with the DBI-action) or the confining string between those quarks. Nonetheless a simple statistical argument shows that our result might be quite plausible. Consider a random distribution of these static $Q\bar{Q}$ pairs in an external magnetic field. Roughly speaking, two thirds of these mesons can be viewed as perpendicular to the applied field, resulting in an increased effective string tension, and hence (using $m^2 \sim \sigma n$) an increased mass of the excitations. Analogously, one third of these mesons are expected to experience a decrease in string tension and mass.

In our case, the different degrees of freedom we have for the mesons are their polarizations, and indeed the two transverse polarizations become more massive, whereas the single longitudinal degree of freedom becomes lighter.

Thus a simple counting argument shows that both results are not inconsistent a priori. A closely connected point was also made in \cite{Bonati:2014ksa}: the chromoelectric field perpendicular to the applied magnetic field increases, whereas the tangential component decreases. Again the direct link between this field and the polarization of the resulting meson is unclear, although it seems plausible that they should be related in some way.

For a recent holographic study on the anisotropy of the heavy quark potential in a magnetic field in the context of $\mathcal{N}=4$ SYM, see \cite{Rougemont:2014efa}.

Obviously a better understanding of this would be beneficial, although this will be very difficult since a clear understanding of the gauge string between quarks and its precise link to the effective meson degrees of freedom is precisely part of the confinement problem.

Unfortunately, we are unaware of any direct experimental or lattice data that can support (or disprove) the above results.

\section{Heavy quark diffusion constant}
Having introduced our model, we are also able to investigate certain transport properties of the magnetized plasma. Here, we shall in particular be interested in the heavy quark diffusion coefficient, which is defined as\footnote{See e.g.~\cite{Bu:2012zza,Bu:2013vk} for other transport related quantities obtained from a different model.}
\begin{equation}
D = \frac{1}{6\chi}\text{lim}_{\omega\to 0}\sum_{i=1}^{3}\frac{\rho^{V}_{ii}}{\omega},
\end{equation}
where the sum is over all three different spatial current-current correlators. In case of a background magnetic field (and hence loss of isotropy), it is natural to define two diffusion coefficients $D_{\perp}$ and $D_{\parallel}$ and it is these quantities that we will study here. The $\chi$ appearing here is the quark number susceptibility. It is expected that this quantity as well will be sensitive to a background magnetic field \cite{Kim:2010zg}. Because $\chi$ is insensitive to the polarization vector (by definition), it will be the same for all polarizations and hence the conclusions below should be interpreted in a relative way: only ratios should be considered.

\subsection{Numerical analysis}
\begin{figure}[h]
\begin{minipage}{0.48\textwidth}
\centering
\includegraphics[width=\textwidth]{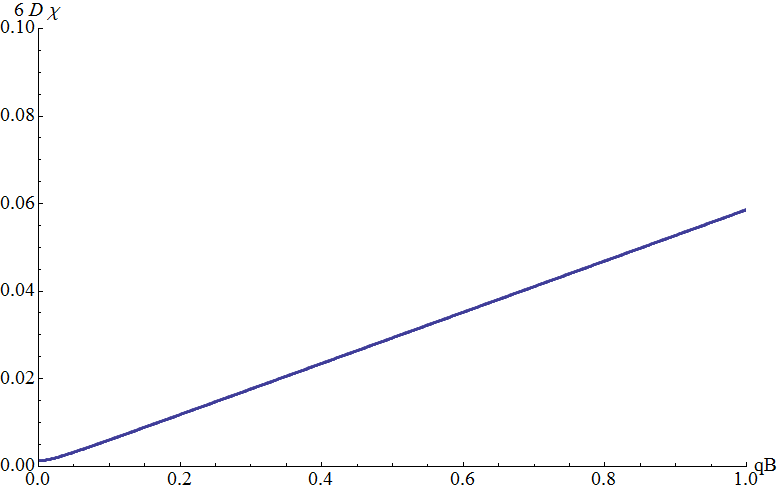}
\end{minipage}
\begin{minipage}{0.48\textwidth}
\centering
\includegraphics[width=\textwidth]{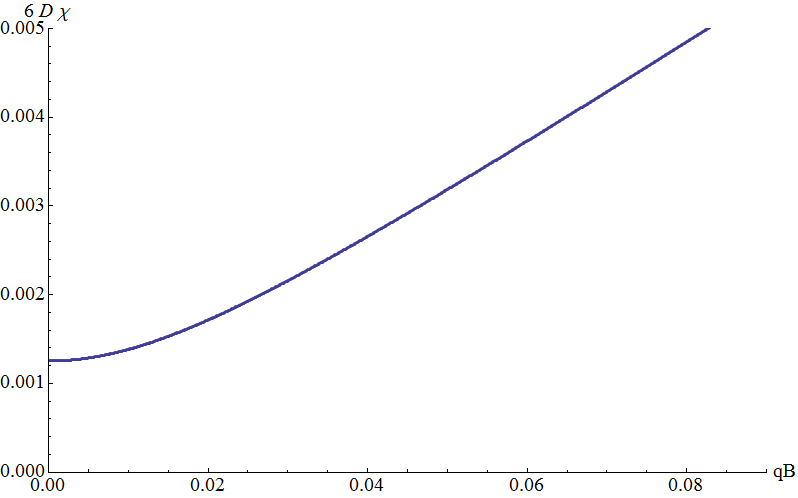}
\end{minipage}
\caption{(a) Heavy quark diffusion coefficient for the polarization parallel to $\mathbf{B}$ as a function of applied magnetic field $qB$ for $t=0.14$. (b) Zoom-in of the previous figure. For $\mathbf{B} \approx \mathbf{0}$, the diffusion constant remains constant. For larger magnetic field, a linear regime sets in.}
\label{diff1}
\end{figure}

\begin{figure}[h]
\centering
\includegraphics[width=0.6\linewidth]{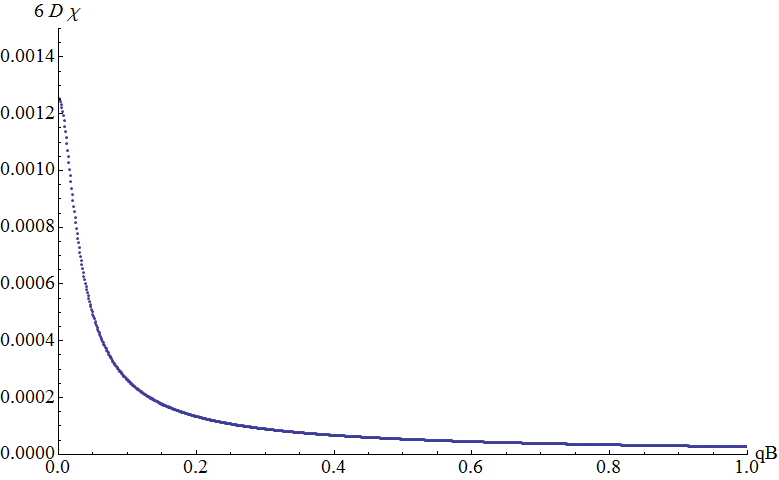}
\caption{Heavy quark diffusion coefficient for the polarization transverse to $\mathbf{B}$ as a function of applied magnetic field $qB$ for $t=0.14$.}
\label{diff3}
\end{figure}

This quantity can be rapidly deduced numerically from the spectral function, and we plot $6\chi D$ as a function of applied magnetic field in FIGS.~\ref{diff1} and \ref{diff3}.


\noindent We observe from these figures that the heavy quark diffusion coefficient for the longitudinal polarization becomes larger than that of the transverse polarizations as the $\mathbf{B}$-field increases. The behavior of $6 D_{\parallel} \chi$ is linear with magnetic field for sufficiently large magnetic fields. As remarked above, a more detailed analysis would require the knowledge of the quark number susceptibility $\chi$ as a function of $\mathbf{B}$ to which we plan to return in future work. \\

\subsection{Diffusion constant using hydrodynamic expansion}
In this subsection we will reproduce the above behavior of $\chi D$ using a hydrodynamic expansion \cite{Son:2002sd,Policastro:2002se}. This will allow us to obtain an analytic formula which provides a check on the numerical work in the previous sections. \\

\noindent Let us return to the differential equation for the longitudinal polarization:
\begin{equation}
V_3'' + \frac{g'}{g}V_3' - \frac{G^{tt}}{G^{33}}\omega^2V_3 = 0,
\end{equation}
where we denoted
\begin{equation}
g = \sqrt{-\mathcal{G}}e^{-cz^2}G^{zz}G^{33} = \sqrt{\frac{L^{10}}{z^{10}} + \frac{L^6}{z^6}D^2}\frac{z^4}{L^4}e^{-cz^2}\left(1-\frac{z^4}{z_h^4}\right)
\end{equation}
for notational simplicity and primes denote $z$-derivatives. The solution $V_3(z)$ is written in the hydrodynamic (small $\omega$) expansion
\begin{equation}
V_3(z)= \left(1-\frac{z}{z_h}\right)^{-\frac{i \omega z_h}{4}}\left(F_0(z) + \omega F_{\omega}(z) + \hdots\right)
\end{equation}
for some unknown functions $F_0(z)$ and $F_\omega(z)$. These functions should not have a singularity at $z=z_h$. Expanding in powers of $\omega$, the lowest order equation is written as
\begin{equation}
(g F_0')' = 0,
\end{equation}
whose general solution is
\begin{equation}
F_0(z) = C_1 \int \frac{dz}{g(z)} + C_2.
\end{equation}
One readily shows that $C_1=0$ to avoid a singularity at $z=z_h$.\footnote{Even though the integral cannot be computed analytically, the divergent part of it can, by extracting the finite parts from the integrand as follows:
\begin{equation}
\int dz f(z) h(z) = f(z_0) \int dz h(z) + \text{finite at $z_0$}
\end{equation}
for a finite function $f$ and a function $h$, singular at $z_0$.
}
The next-to-lowest order equation in $\omega$ yields, after a first integration,
\begin{equation}
\left(\frac{i\omega}{4}\left(1-\frac{z}{z_h}\right)^{-\frac{i \omega z_h}{4}-1}C_2 + \left(1-\frac{z}{z_h}\right)^{-\frac{i \omega z_h}{4}}\omega F'_{\omega}\right) = \frac{C_3}{g},
\end{equation}
leading to
\begin{equation}
F'_{\omega} = \frac{C_3}{\omega g}\underbrace{\left(1-\frac{z}{z_h}\right)^{\frac{i \omega z_h}{4}}}_{=1+\mathcal{O}(\omega)} - \frac{i C_2}{4\left(1-\frac{z}{z_h}\right)}.
\end{equation}
Note that this equation implies that $C_3 =\mathcal{O}(\omega)$ by definition of our expansion. This leads to
\begin{equation}
F_{\omega} = \int \frac{C_3 dz}{\omega g} + \frac{i C_2 z_h}{4}\ln\left(1-\frac{z}{z_h}\right) + C_4.
\end{equation}
The second term has a singularity at $z=z_h$, but the first term has one as well. If one chooses
\begin{equation}
\frac{C_3}{\omega}\frac{Lz_h^2 e^{cz_h^2}}{4\sqrt{L^4+D^2z_h^4}} = \frac{iC_2 z_h}{4},
\end{equation}
the singularity vanishes. Finally $C_4$ needs to be chosen such that $F_{\omega}(z=z_h)$ vanishes, but its precise value will not be needed. Combining these results, the expansion becomes
\begin{equation}
V_3(z)= C_2\left(1-\frac{z}{z_h}\right)^{-\frac{i \omega z_h}{4}}\left(1 + \frac{i\omega e^{-cz_h^2}}{z_h L}\sqrt{L^4+D^2z_h^4}\int_{z^*}^{z}\frac{dz}{g(z)} + \frac{i\omega z_h}{4}\ln\left(1-\frac{z}{z_h}\right)\right),
\end{equation}
where $z^*$ incorporates $C_4$ and is chosen such that $F_{\omega}(z=z_h)=0$. To get the retarded Green function, one needs $\frac{V_3 \partial_z V_3}{z}$ in the $z\to 0 $ limit. Moreover, to get $\chi D$ one should divide this expression by $\omega$ and take $\omega \to 0$. \\
Firstly, the boundary value of $V_3$ is given by
\begin{equation}
V_3(z=0) = C_2\left(1 + \frac{i\omega e^{-cz_h^2}}{z_h L}\sqrt{L^4+D^2z_h^4}\int_{z^*}^{0}\frac{dz}{g(z)}\right),
\end{equation}
which determines $C_2$. We can then write
\begin{align}
\partial_z V_3 &= C_2\frac{i\omega}{4}\left(1-\frac{z}{z_h}\right)^{-\frac{i \omega z_h}{4}-1}\left(1 + \frac{i\omega e^{-cz_h^2}}{z_h L}\sqrt{L^4+D^2z_h^4}\int_{z^*}^{z}\frac{dz}{g(z)} + \frac{i\omega z_h}{4}\ln\left(1-\frac{z}{z_h}\right)\right) \\ \nonumber
&+ C_2\left(1-\frac{z}{z_h}\right)^{-\frac{i \omega z_h}{4}}\left( \frac{i\omega e^{-cz_h^2}}{z_h L}\sqrt{L^4+D^2z_h^4}\frac{1}{g(z)} - \frac{i\omega}{4}\frac{1}{1-z/z_h}\right).
\end{align}
Dropping the term quadratic in $\omega$ and taking $z\to 0$, this reduces to
\begin{equation}
\partial_z V_3(z=0) = C_2\left( \frac{i\omega e^{-cz_h^2}}{z_h L}\sqrt{L^4+D^2z_h^4}\frac{1}{g(0)}\right) + \mathcal{O}(\omega^2).
\end{equation}
We hence finally obtain for $\chi D_{\parallel}$:
\begin{equation}
\chi D_{\parallel} \sim \Im\text{lim}_{\omega\to 0} \frac{1}{\omega}\text{lim}_{z\to 0} \frac{V_3 \partial_z V_3}{z} \sim V_3(z=0)C_2\left(\frac{ e^{-cz_h^2}}{z_h L}\sqrt{L^4+D^2z_h^4}\frac{1}{L}\right) \sim \sqrt{L^4+D^2z_h^4},
\end{equation}
where we used $g(0) = \frac{L}{z}$ and the low $\omega$ limit implies also that $V_3(z=0) \approx C_2$. Note that since $D^2 \sim L^4$, the result is independent of $L$, as has been argued for previously.\\

\begin{figure}[h]
\centering
\includegraphics[width=0.6\textwidth]{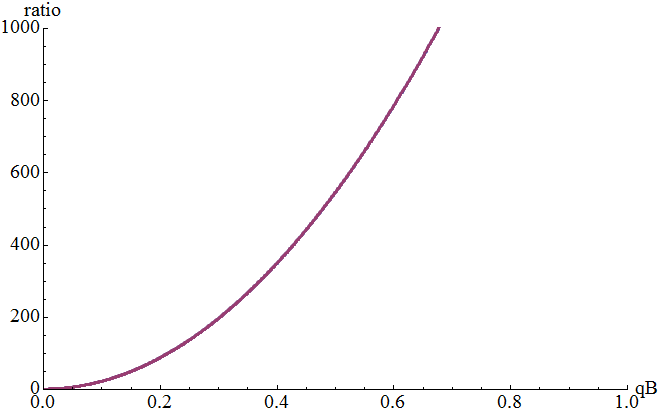}
\caption{The ratio $\frac{D_{\parallel}}{D_{\perp}}$ computed both numerically and using the analytic formula (for $t=0.14$). Both results coincide.}
\label{ratio}
\end{figure}

\noindent Taking transverse polarizations entails an almost identical computation. The only difference is the value of $C_3$ in terms of $C_2$ to avoid a singularity at $z = z_h$:
\begin{equation}
\frac{C_3}{\omega}\frac{z_h^2 e^{cz_h^2}\sqrt{L^4+D^2z_h^4}}{4L^3} = \frac{iC_2 z_h}{4}.
\end{equation}
In the end this simply leads to
\begin{equation}
\sqrt{L^4+D^2z_h^4} \to \frac{L^2}{\sqrt{L^4+D^2z_h^4}}.
\end{equation}
Ignoring some prefactors, the final result we obtain is
\begin{align}
6\chi D_{\parallel} &\sim \frac{e^{-cz_h^2}}{z_h \pi}\sqrt{1 + \frac{16\pi^2(2.29)^4z_h^4}{9}(qB)^2}, \\
6\chi D_{\perp} &\sim \frac{e^{-cz_h^2}}{z_h \pi}\frac{1}{\sqrt{1 + \frac{16\pi^2(2.29)^4z_h^4}{9}(qB)^2}}.
\end{align}
The ratio of longitudinal and transverse $D$ becomes
\begin{equation}
\frac{D_{\parallel}}{D_{\perp}} = \frac{L^4+D^2z_h^4}{ L^4} = 1 + \frac{16\pi^2(2.29)^4z_h^4}{9}(qB)^2.
\end{equation}
This analytic formula and the one obtained numerically agree and both are plotted in FIG.~\ref{ratio}.
It is clear that both methods yield the same outcome, supporting both our numerical procedure and the above analytic results. \\
We find it quite remarkable that the final results can be written down analytically even though the integration over $1/g(z)$ cannot be done analytically.

\section{Conclusion and outlook}
Summarizing, in this paper we have introduced an extension of the original soft wall AdS/QCD model that is capable of describing the coupling of an external magnetic field to the charged constituents of charmonia. We employed the model to probe the melting properties of the $J/\psi$ in the presence of a magnetic field. Our main finding is that the polarizations perpendicular, resp.~parallel, to the applied magnetic field melt faster, resp.~slower than in the case where no magnetic field is present. As expected, the anisotropy brought in by the magnetic field also induces an anisotropy in the melting of the heavy vector meson. It is beyond doubt clear that the magnetic field does influence the melting, and as such, it gives support to the picture that the deconfinement transition depends on the applied magnetic field, although in a more involved way than the Polyakov loop picture might suggest \cite{Bali:2011qj}. Let us mention here that we did not need to make the soft wall background itself dependent on the magnetic field to find a nontrivial effect. Although our analysis might seem to correspond to a probe brane approximation, this is actually to be loosely interpreted, since it would be useless to consider the backreaction of the magnetic field on the soft wall metric, since the latter does not solve itself Einstein's equations.

As a further application, one might want to look at a background electric field. However, a moment's thought shows that this immediately leads to an imaginary part in the action, no matter how small the applied $\mathbf{E}$ field. In general, one expects a Schwinger-type effect to take place only as soon as the confining force is conquered by the applied field \cite{Sato:2013dwa,Kawai:2013xya,Hashimoto:2014dza}. The soft-wall DBI model is hence not capable of predicting this expected phenomenon correctly.\footnote{This could be implemented by using a different pull-back metric in the DBI-action, but doing this would compromise our wish to reduce to the normal soft-wall action for vanishing electromagnetic background.} In general, soft-wall type models appear to be less powerful in predicting properties directly related to the confining force (Wilson loop with area law,\footnote{For a discussion on this, see \cite{Karch:2010eg}. Since the soft wall model cannot display the linear potential, there was a priori little hope to recover a phenomenologically viable Schwinger effect in a confining theory.} electric breakdown etc.), but its strength lies in the prediction of the (linear) meson Regge trajectory, and it is our hope that these poles have been qualitatively predicted using our DBI-modification of this model.

We also presented a preliminary analysis of the relative heavy quark diffusion constants for longitudinal and transversal modes. The former modes display a stronger diffusion than the latter ones. It is tempting to speculate that this observation might have interesting consequences for the elliptic flow in the medium \cite{Petreczky:2005nh}, which is anyhow related to anisotropy. Since the collision process creates the magnetic field, another proponent of anisotropy as we have confirmed e.g.~in the melting or transport properties of our model, it looks worthwhile to pursue this in more detail. This will encompass the effective computation of the heavy quark number susceptibility.

\section*{Acknowledgments}
We thank D.~R.~Granado for a careful reading of the manuscript. The work of T.~G.~Mertens is supported by the Special Research Fund of Ghent University.

\end{document}